\documentclass[pdflatex,sn-mathphys-num]{sn-jnl}

\usepackage{graphicx}%
\usepackage{multirow}%
\usepackage{amsmath,amssymb,amsfonts}%
\usepackage{amsthm}%
\usepackage{mathrsfs}%
\usepackage[title]{appendix}%
\usepackage{xcolor}%
\usepackage{textcomp}%
\usepackage{manyfoot}%
\usepackage{booktabs}%
\usepackage{algorithm}%
\usepackage{algorithmicx}%
\usepackage{algpseudocode}%
\usepackage{listings}%

\theoremstyle{thmstyleone}%
\theoremstyle{thmstyletwo}%

\theoremstyle{thmstylethree}%

\usepackage{amsmath,amsfonts,bm}

\def\eqref#1{equation~\ref{#1}}

\def\1{\bm{1}}

\DeclareMathAlphabet{\mathsfit}{\encodingdefault}{\sfdefault}{m}{sl}
\SetMathAlphabet{\mathsfit}{bold}{\encodingdefault}{\sfdefault}{bx}{n}

\usepackage{hyperref}
\usepackage{url}
\usepackage{caption}
\DeclareCaptionLabelFormat{myformat}{\textbf{#1 #2 \textbar}}
\title{A Universal Space of Brain Dynamics for Unveiling Cognitive Transitions and Individual Differences}

\author[1]{\fnm{Ronghua} \sur{Zheng}} 
\author[1]{\fnm{Chengyuan} \sur{Qian}}
\author*[1,2]{\fnm{Weiyang} \sur{Ding} \email{dingwy@fudan.edu.cn}}

\affil[1]{\orgdiv{Institute of Science and Technology for Brain-Inspired Intelligence}, \orgname{Fudan University}, \orgaddress{\city{Shanghai}, \country{China}}}
\affil[2]{\orgdiv{Key Laboratory of Computational Neuroscience and Brain-Inspired Intelligence (Fudan University)}, \orgname{Ministry of Education}, \orgaddress{\country{China}}}

\begin{document}

\maketitle

\begin{abstract}

Representing dynamical systems through data-driven universal spaces has proven effective; however, achieving this universality for human brain activity remains a significant challenge, further aggravated by diverse cognitive states and individual subjects. Recognizing that spatial properties reflect physical wiring while temporal properties reflect brain function, we develop Universal Brain Dynamics (UBD) to construct a universal space tailored to brain activity and quantify corresponding dynamics using a model-derived Jacobian matrix. Crucially, we validate UBD’s universality by accurately predicting functional magnetic resonance imaging (fMRI) signals (Pearson's $r>0.9$) across eight states and 963 subjects in the Human Connectome Project (HCP). Through evaluating resting-state fMRI represented within UBD, we gain insight into how infra-slow fluctuation (ISF) underpins brain activity. Furthermore, we reveal a new perspective on structure-function coupling (SFC) by analyzing the temporal sequence of brain dynamics. Extending UBD to task-evoked states, we derive brain dynamics across various cognitive conditions, elucidating the neural mechanisms driving cognitive transitions at a finer granularity. For individual differences, we compare brain dynamics across subjects to identify the neural underpinnings of these variations. Our findings suggest that synergistically integrating spatial and temporal properties of brain activity establishes a universal space for its unfolding, enabling the precise numerical analysis of underlying neural mechanisms across varying conditions.

\end{abstract}

\section{Introduction}
The study of human brain activity is an inexhaustible landscape, rooted in the unparalleled complexity of neural matter, unfolding across diverse scenarios, and ultimately modulated by the unique signature of the individual mind. The infinite conditions do not imply that the principles governing them are equally fragmented. Consider a symphony orchestra concert: listeners of varying genders, ages, and backgrounds can follow the conductor and resonate with the spirit the composer seeks to convey. This phenomenon suggests that diverse patterns of brain activity can unfold within a shared, underlying space (e.g., manifold \citep{shine2019human,chaudhuri2019intrinsic}).
Recent research has seen a proliferation of studies demonstrating that unified frameworks can successfully extract common representations of brain activity from heterogeneous perspectives \citep{pang2023geometric,raut2025arousal,beaglehole2026toward}. This evidence supports a compelling hypothesis: that diverse neural activities are grounded in a universal latent space \citep{raut2025arousal}. But does this universality hold when accounting for diverse cognition and distinct individuals?

Theoretically, brain activity can be conceptualized as coordinated dynamics evolving within a unified system, where specific properties act as constraints or boundary conditions \citep{deco2015rethinking,breakspear2017dynamic,gosztolai2025marble}. Yet, the methodology to explicitly identify this system and analyze the dynamics within it remains elusive.
Deep learning has demonstrated remarkable success in representing non-linear systems by constructing data-driven latent spaces \citep{brunton2016discovering}, a strength further enhanced by the Koopman theory of dynamical systems \citep{brunton2017chaos,luschDeepLearningUniversal2018}. 
The central focus of modern Koopman analysis is to find nonlinear coordinate transformations that linearize dynamics \citep{brunton2022modern}; however, identifying the span of these transformations to accurately represent complex brain activity under varying conditions remains a challenge. To overcome these limitations, we develop a deep learning framework that simultaneously accounts for both the spatial and temporal properties of brain activity. In general, the spatial property is delineated by structural connectivity (SC) \citep{yeh2021mapping}, depicting the physical wiring of the brain; the temporal property is captured by functional connectivity (FC), reflecting statistical dependencies of brain function \citep{van2010exploring}. Therefore, synergistically integrating these two properties is essential for establishing a universal latent space tailored to brain activity \citep{park2013structural,suarez2020linking}.

Having established a universal latent space that precisely represents brain activity, the subsequent challenge lies in interpreting the underlying neural mechanisms. By conceptualizing brain areas as nodes and their connections as edges, graph-based methods (e.g., spectral decomposition, network diffusion) provide a powerful methodology for both representing and explaining brain activity \citep{pang2023geometric,yang2023enhanced,bassett2017network}. Specifically, graph convolutional networks (GCNs) excel at modeling indirect, higher-order nodal connections \citep{kipf2017semi}, making them ideally suited for capturing the multi-synaptic interactions across the brain \citep{goni2014resting}. Since the routing of these interactions is physically constrained by the brain's anatomical network \citep{avena2018communication}, we use SC to guide the GCN's message passing across all brain areas.

The inherent "black-box" nature of neural networks hinders the interpretability of GCNs \citep{lipton2018mythos,montavon2018methods}. To circumvent this problem, we employ the Jacobian matrix to reveal the underlying mechanisms of brain activity \citep{baydin2018automatic,luo2025mapping}. Mathematically, the Jacobian matrix captures the time-varying relationships between inputs and outputs, providing a rigorous numerical tool to quantify the dynamics of brain activity. When combined with the aforementioned universal space, this paradigm enables us to evaluate model-derived dynamics across diverse cognitive states and subject cohorts, revealing the neural mechanisms driving cognitive transitions and the neural underpinnings of individual differences.

Here, we propose a paradigm to interpret brain activity across varying cognitive and individual conditions via model-derived dynamics. To this end, we developed Universal Brain Dynamics (UBD) to represent fMRI signals by embedding diffusion-weighted magnetic resonance imaging (dMRI) into the neural network's backbone and utilizing temporal fMRI data for training, thereby mimicking the brain's intrinsic wiring and function. We validate the universality of UBD by demonstrating its unprecedented predictive accuracy and generalizability. By explicitly representing fMRI signals within UBD, we gain deeper insights into how infra-slow fluctuation (ISF) underpins brain activity \citep{zuo2010oscillating}. Furthermore, we characterize brain dynamics using a GCN-derived Jacobian matrix, establishing a novel perspective on structure-function coupling (SFC) \citep{fotiadis2024structure}. Crucially, the predictive performance of UBD enables us to derive brain dynamics across multiple cognitive states and subjects, thereby elucidating the mechanisms underlying cognitive transitions and individual differences. Our findings suggest that synergistically integrating the spatial and temporal properties establishes a universal space of brain dynamics, enabling the precise numerical analysis of complex brain functions.

\section{Results}

\subsection{Development and validation of UBD}
\begin{figure}[htbp]
	\centering
	\includegraphics[scale=0.721]{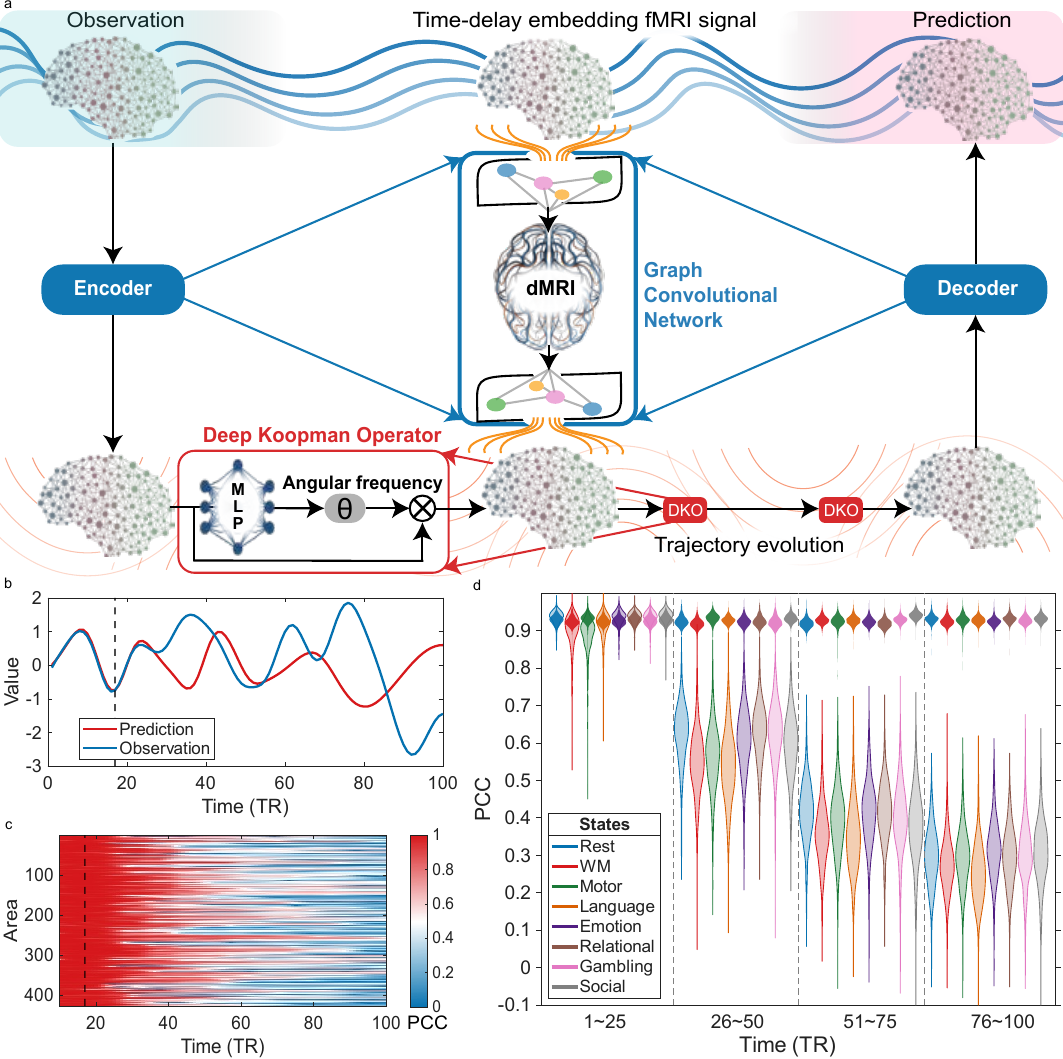}
	\caption{\textbf{Framework architecture and prediction results.} \textbf{a}, overview of UBD. The top panel shows a time-delay embedding fMRI signal in the measurement space, while the bottom panel depicts the corresponding trajectory in the universal latent space. 
    The prediction pipeline consists of three main stages: an encoder that maps an observed fMRI snapshot to a latent representation, a DKO that predicts subsequent latent representations, and a decoder that maps them back into predicted fMRI signals. The encoder is implemented as a GCN, where brain areas are treated as nodes and time-delay embeddings as features, with message passing guided by dMRI. The decoder mirrors the encoder architecture and preserves the dimensionality of the fMRI snapshots, enabling a seamless mapping back to the measurement space. The DKO employs an MLP to estimate the angular frequency $\bm{\theta}$ from the current latent representation and applies complex multiplication to generate the next representation. Recursively applying this process captures the evolution of the latent trajectory and enables multi-step prediction.
    \textbf{b}, prediction results (time length=100) for a single brain area. Red and blue curves indicate predicted and observed signals, respectively. The vertical dashed line demarcates the initialization window used for time-delay embedding. \textbf{c}, prediction results for a single subject across 426 brain areas. Each row illustrates the temporal variation of the PCC between prediction and observation in a specific area. \textbf{d}, distribution of the PCC across 963 subjects and 8 cognitive states (color-coded). Dashed lines divide the prediction horizon into four intervals. Each violin plot illustrates the distribution of group PCC within the corresponding interval. Translucent plots represent multi-step predictions generated from initial data, while bold plots represent one-step predictions based on immediately preceding data.}
    \label{pic:1}
\end{figure}

Establishing a universal framework for brain activity necessitates following its essential temporal and spatial properties, which are constrained by sensing technology and brain anatomy.
We illustrate the overview of UBD in Fig.\hyperref[pic:1]{1a}. In the measurement space (top panel), we apply Takens' theory \citep{takens2006detecting} to stack fMRI signals into snapshots, thereby overcoming their inherent limitations in temporal resolution. An encoder GCN conceptualizes brain areas as graph nodes, with their time-delayed information forming nodal features \citep{mitra2015lag}. 
Guided by dMRI-derived SC routing, the GCN propagates features across nodes to map fMRI snapshots into latent representations (bottom panel). To reflect temporal evolution, UBD employs a deep Koopman operator (DKO) to learn the angular frequency $\bm{\theta}$ between consecutive latent representations via complex multiplication. Finally, a decoder GCN with the identical SC maps the evolved latent representations back to the observation space to predict fMRI signals.
We derive two loss functions from this framework: a prediction loss, calculated by comparing observed and predicted fMRI snapshots; and a latent loss, computed by comparing latent trajectories projected by the DKO from the initial time against trajectories encoded by the GCN at each time step. The prediction loss ensures accurate fMRI predictions, while the latent loss enforces consistency between these signals and latent trajectories. This combined loss was used to train the model on resting-state fMRI data from 35 subjects (Methods). We derived all subsequent results from this single trained model.

We showcased the universality of UBD by testing its prediction accuracy and generalization performance across various cognitive states and subjects. For clarity, we first illustrated predicted and observed data for a single brain area (Fig.\hyperref[pic:1]{1b}, prediction horizon=100), then expanded this result to the whole brain by computing the Pearson correlation coefficient (PCC) between predictions and observations across brain areas (Fig.\hyperref[pic:1]{1c}). The PCC neared 1 until $t=20$ across all areas and its average sustained approximately 0.5 up to $t=50$, thereby confirming UBD's capability to predict whole-brain signals precisely. Subsequently, we extended this analysis from single-subject resting-state data to several task-evoked states across a broader cohort. By dividing the prediction horizon into four intervals, we demonstrated the PCC distribution for 963 subjects within their corresponding intervals (Fig.\hyperref[pic:1]{1d}, eight states denoted by colors). During the initial interval ($t=1$--25), the mean PCC robustly achieved 0.9 irrespective of the cognitive state. As time progresses to the final interval ($t=7$6--100), the accuracy for the resting-state (blue) settles at 0.3, with performance in the other states tracking slightly below this level. The bold violin plots represent one-step prediction results, which in turn signify that the transparent violin plots are replicable for the next 100 time points. We further evaluated prediction accuracy using resting-state fMRI from 100 UK Biobank (UKB\citep{sudlow2015uk,miller2016multimodal,alfaro2018image}, Extended Data Fig.\ref{pic:sup_ukb}) subjects; The comparable performance observed between HCP and UKB suggests that UBD generalizes well across datasets. Remarkably, despite training UBD on resting-state data from only 35 subjects, the prediction results remained robust and replicable across all subjects and states. This consistent prediction accuracy across time, states, subjects, and datasets strongly suggests that UBD creates a universal latent space well-tailored to brain activity, thereby forming the cornerstone for subsequent analyses.

\subsection{Exploring resting-state fMRI signals within UBD}
\label{rest}
\begin{figure}[htbp]
	\centering
	\includegraphics[scale=0.725]{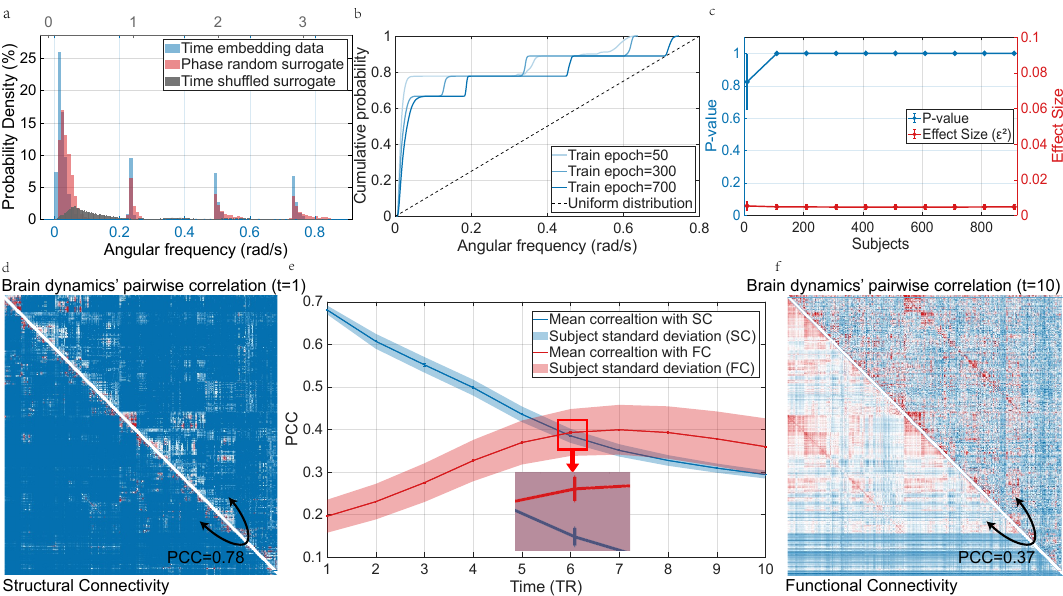}
	\caption{\textbf{Spatiotemporal properties of resting-state fMRI signal revealed by UBD.} \textbf{a}, distribution of the angular frequency $\bm{\theta}$ of a resting-state fMRI signal (time length=10). The blue histogram peaks span 0--0.09, 0.22--0.25, 0.49--0.53, and 0.72--0.78 rad/s, which correspond to frequency bands of 0--0.014, 0.035--0.040, 0.078--0.084, and 0.115--0.124 Hz, respectively. The red histogram shows $\bm{\theta}$ from phase-randomized surrogates and shares the x-axis with the blue histogram. The gray histogram shows $\bm{\theta}$ from time-shuffled surrogates, with its range indicated on the top gray x-axis.     
    \textbf{b}, cumulative distribution function (CDF) of $\bm{\theta}$. The bold blue curve (Train epoch=700) indicates that 0–0.09 rad/s dominates (around 65$\%$), while 0.22–0.25, 0.49–0.53, and 0.72–0.78 rad/s each contribute around 10$\%$. Translucent blue curves show $\bm{\theta}$ from the same fMRI data across different training epochs. The diagonal dashed line represents a uniform distribution benchmark.    
    \textbf{c}, the consistent distribution of $\bm{\theta}$  across subjects ($n=963$). The blue and red curves illustrate the P-value and effect size  ($\varepsilon^{2}$) derived from the Kruskal-Wallis test. As the sample size increases, the P-value converges to 1 and the effect size remains negligible ($\sim$0).    
    \textbf{d}, comparison of SC and brain dynamics' pairwise correlation at $t=1$. The PCC was calculated between the SC used to train UBD and the pairwise correlations, then averaged across 963 subjects.
    \textbf{e}, relationship between brain dynamics' pairwise correlation with SC and FC. The blue curve shows a decreasing correlation with SC over time, while the red curve shows an increasing correlation with FC. Shaded areas indicate the standard deviation across subjects, and inset error bars show the standard deviation across time.
    \textbf{f}, comparison of FC and brain dynamics' pairwise correlation at $t=10$. The PCC was calculated between each subject's FC and their corresponding pairwise correlation, then averaged across 963 subjects.}
    \label{pic:2}
\end{figure}

Since intractable fMRI signals can be accurately represented in the universal latent space established by UBD, we can analyze them through the corresponding latent trajectories.
As the evolution of a latent trajectory is determined by the angular frequency $\bm{\theta}$ generated by the DKO, the temporal properties of the original fMRI signal can be characterized through $\bm{\theta}$. Fig.\hyperref[pic:2]{2a} illustrates the distribution of $\bm{\theta}$ for a resting-state fMRI signal (time length=10), showing four prominent peaks (blue histogram) within the ranges 0--0.09, 0.22--0.25, 0.49--0.53, and 0.72--0.78 rad/s. According to the relation $f=\bm{\theta}/(2\pi)$ (Methods), these angular frequency ranges correspond to frequency bands of 0--0.014, 0.035--0.040, 0.078--0.084, and 0.115--0.124 Hz, respectively. 
To confirm the peaks phenomenon reflects intrinsic fMRI signal properties, we analyzed phase-randomized (red histogram) and time-shuffled (gray histogram) surrogate data. The phenomenon's persistence in phase-randomized surrogates and absence in time-shuffled surrogates demonstrates its dependence on the signal's frequency profile. Additional comparisons of the distribution of $\bm{\theta}$ and the original fMRI signal’s amplitude spectrum (Extended Data Fig.\ref{pic:sup_spectrum}) further support this conclusion.
Fig.\hyperref[pic:2]{2b} illustrates the cumulative density function (CDF) of the distribution of $\bm{\theta}$; the lowest-value peak is dominant, while the remaining components contribute nearly equally (bold blue curve, train epoch=700). By comparing this CDF to those generated at earlier training epochs (translucent blue lines), we observed that these peaks became sharper as training progressed. 
We utilized the Kruskal-Wallis (KW) test to demonstrate that this peak distribution of $\bm{\theta}$ is highly consistent across subjects (Methods); Fig.\hyperref[pic:2]{2c} shows that the P-values (blue curve) and effect sizes ($\epsilon^{2}$, red curve) stabilize as the number of subjects increases to 963.
Overall, these results indicate that resting-state fMRI signals are driven by four distinct oscillatory components that align with classically defined frequency bands (e.g., Slow-5, 0.01--0.027 Hz; Slow-4, 0.027--0.073 Hz). Notably, the lowest-frequency yet dominant component likely corresponds to the Slow-5 band, which primarily captures large-scale cortical activity, whereas the second lowest-frequency peak may reflect the Slow-4 band, associated with more localized subcortical and sensorimotor processes \citep{zuo2010oscillating}.

We further characterized spatiotemporal brain dynamics via the GCN-derived Jacobian matrix (Methods). Beyond mapping fMRI signals to latent trajectories, the GCN uses its Jacobian matrix to quantify the influence of each brain area: each column captures an area’s effect on the latent trajectories, and the full matrix represents whole-brain influence. Extending this analysis to a temporal sequence of Jacobian matrices reveals how this influence evolves, thereby quantifying the dynamics of brain activity. 
Following the concept of FC, which assesses similarity among areas, we computed the pairwise correlations of brain dynamics for resting-state fMRI signals in 963 subjects (time length=10). Fig.\hyperref[pic:2]{2d} demonstrates a strong similarity between SC and this pairwise correlation at $t=1$. This stands in stark contrast to Fig.\hyperref[pic:2]{2f}, which highlights the alignment between FC and the pairwise correlation at $t=10$. Fig.\hyperref[pic:2]{2e} quantifies this transition: the PCC between the pairwise correlation and SC (blue curve) starts high ($\geq 0.75$) and declines, while the PCC with FC (red curve) shows an inverse trend, exceeding the blue curve at $t=5$ and achieving around 0.4 at $t=7$. To verify that these findings arise from the synergistic integration of spatial and temporal characteristics of brain activity, we replaced SC with a random matrix (Extended Data Fig.\ref{pic:sup_random}), which completely abolished the observed effects. Collectively, brain dynamics derived from UBD offer a novel perspective, reframing SFC not as a static association but as an evolving process where SC serves as the origin and FC as the destination.

\subsection{Dissecting cognitive transition via brain dynamics}
\label{task}
\begin{figure}[htbp]
	\centering
	\includegraphics[scale=0.725]{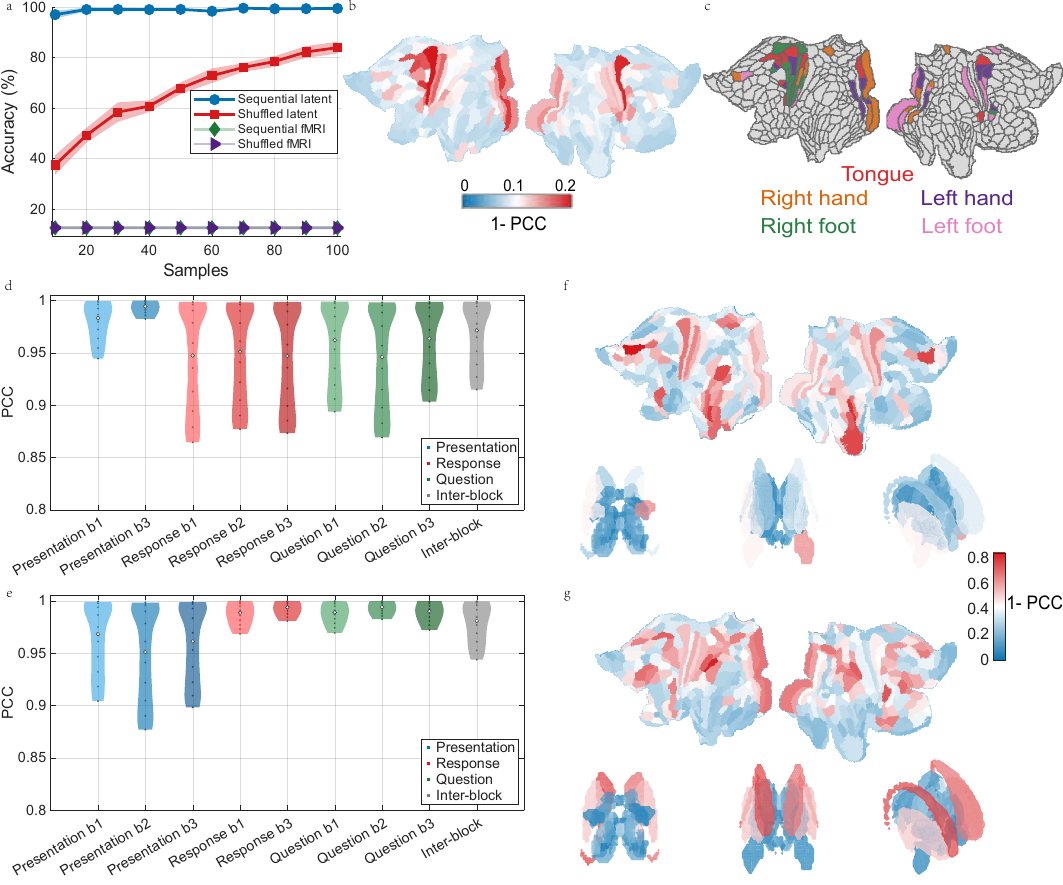}
	\caption{\textbf{Quantifying cognitive transitions and divergence through brain dynamics.}
    \textbf{a}, classification accuracy for eight states using four data types. Shuffled latent data (red curve) shows accuracy rising from 40\% to 80\% with increasing sample size (from 10 to 100). Shaded areas indicate the standard deviation across subjects. Sequential latent data (blue curve) achieves nearly 100\% accuracy regardless of sample size.  In contrast, both sequential and shuffled fMRI data (green and purple curves) remain at chance level  (12.5\%).     
    \textbf{b}, dissimilarity between brain dynamics derived from the tongue and fixation conditions at $t=5$ ($n=915$).
    \textbf{c}, the contrast map highlights the top 10 dissimilar areas across five movements (e.g., tongue movement in \textbf{b}). Brain region boundaries are outlined, and multi-colored areas indicate the corresponding movements.
    \textbf{d--e}, relationship of brain dynamics across conditions and time blocks within the language task. Benchmarks are defined as the brain dynamics derived from the presentation and response conditions at time block 2 (time length=10, $n=920$). Panels \textbf{d} and \textbf{e} show the similarity of the remaining conditions (color-coded) to the presentation and response benchmarks, respectively. The x-axis labels (e.g., “presentation b1”, “question b3”) denote condition and time block, where the letter indicates the condition and the number indicates the corresponding time block. 
    \textbf{f--g}, divergence of brain dynamics between the presentation and response conditions. Panels \textbf{f} and \textbf{g} display dissimilarity between brain dynamics derived from the inter-block period and those of the presentation and response benchmarks at $t=10$.}
    \label{pic:3}   
\end{figure}

Building on the insight of brain dynamics in resting-state, we explore the mechanisms driving cognitive transitions during task-evoked states. To demonstrate the discriminative capacity of brain dynamics, we classified latent trajectories sampled from 20 subjects into eight cognitive states and compared the accuracy with that of the corresponding raw fMRI signals (Methods). As shown in Fig.\hyperref[pic:3]{3a}, the classification accuracy achieved using sequential latent trajectories (blue curve) approaches 100\%, independent of sample size. To isolate the contribution of the GCN, we temporally shuffled the latent trajectories before classification. The accuracy using these shuffled trajectories (red curve) increases from 40\% to 80\% as sample size grows, indicating that the GCN encodes substantial task-related information even in the absence of temporal structure. In contrast, the results using sequential and shuffled fMRI signals (green and purple lines) remain at chance-level (12.5\%). Together, these results demonstrate that task-related information is far more distinguishable in the universal latent space than in the original measurement space, confirming that GCN-derived brain dynamics capture essential discriminative features across task conditions.

To identify the mechanisms driving cognitive transitions, we compared brain dynamics across conditions using dynamics correlation (DC, Methods). Given that the five movements within the motor task are well-studied \citep{penfield1937somatic}, we derived brain dynamics for the fixation condition and five movement conditions in 915 subjects \citep{barch2013function}. To quantify the divergence across movement conditions, we calculated DC between the fixation condition and each movement condition. We visualized $1-$DC to highlight the dissimilarity between the tongue and fixation condition (Fig.\hyperref[pic:3]{3b}) and map contrasts across the five movements (Fig.\hyperref[pic:3]{3c}). Dissimilarities between the fixation and remaining four movement conditions are shown in Extended Data Fig.\ref{pic:sup_mot}.
We observed a sensorimotor–control network with effector-specific topography, aligning with multimodal parcellations \citep{glasser2016multi}. All movements engaged contralateral \citep{penfield1950cerebral} and weaker ipsilateral \citep{kim1993functional,diedrichsen2013two} primary motor and somatosensory cortices, alongside premotor, supplementary motor, and dorsal mid-cingulate regions for cognitive–motor integration \citep{shackman2011integration}. Visual cues additionally recruited visual, frontal eye fields, and posterior parietal visuomotor and attention systems \citep{corbetta2002control,culham2006human}, plus frontoparietal and cingulo-opercular top-down control networks \citep{dosenbach2008dual,cole2007cognitive}. Within this shared architecture, effector-specific specializations were evident: tongue movements engaged the ventrolateral orofacial regions, frontal opercular, insular, and parietal opercular regions \citep{brown2008larynx,grabski2012functional}; hand movements emphasized lateral premotor and dorsal premotor–parietal circuits, including inferior parietal regions associated with visually guided reaching and hand–eye coordination \citep{grefkes2004human,filimon2007human}; and foot movements recruited the paracentral lobule and supplementary and superior parietal regions \citep{sahyoun2004towards,kapreli2007lower}.

We further demonstrate that cognitive transitions align more closely with specific cognitive conditions than with chronological time within the language task. Specifically, we derived brain dynamics during the presentation, question, and response conditions (time length=10) across three time blocks in 920 subjects. For comparative analysis, we selected brain dynamics from the presentation and response conditions in time block 2 as benchmarks, and computed DC between these benchmarks and the remaining conditions. Fig.\hyperref[pic:3]{3d} shows greater DC between the presentation benchmark and presentation condition across time blocks (blue) than between different conditions within the same block. Fig.\hyperref[pic:3]{3e} reinforces these primary conclusions, while high DC between the response benchmark and the question condition may stem from the short time window and time-delay embedding. Notably, DC between the benchmarks and the inter-block period (gray) is second only to that between identical conditions, indicating that brain dynamics differ more between distinct cognitive conditions than between a cognitive condition and the inter-block period. To further illustrate this divergence, we compare $1-$DC between the presentation and response conditions with the inter-block period (Supplementary Videos 1 and 2), plotting the process at t=10 in Fig.\hyperref[pic:3]{3f--g}. Fig.\hyperref[pic:3]{3f} shows a pattern consistent with distributed accounts of semantic processing \citep{binder2009semantic}, engaging a bilateral temporal–frontal network including superior and middle temporal cortex, anterior temporal regions, and left inferior frontal cortex \citep{hickok2007cortical,friederici2011brain}, supporting lexical–semantic access and combinatorial processing. Medial prefrontal and posterior midline regions (posterior cingulate cortex, precuneus) are also recruited, consistent with integrative and situation-level processing \citep{buckner2008brain,spreng2009common,mar2011neural}. Early auditory and visual cortices are engaged, consistent with ongoing perceptual processing \citep{felleman1991distributed,kayser2007functional}. Activity in the amygdala and ventral/dorsal striatum is consistent with affective or motivational relevance \citep{lindquist2012brain,haber2010reward}, and sensorimotor cortices are engaged in line with partial embodiment accounts \citep{hauk2004somatotopic}.
Fig.\hyperref[pic:3]{3g} shows that during the response phase, the pattern shifts to dorsal frontoparietal and cingulo-opercular systems, including superior parietal cortex, intraparietal sulcus, frontal eye fields, and premotor/motor regions, consistent with visuospatial attention, oculomotor control, and task maintenance \citep{corbetta2002control,seeley2007dissociable}. Dorsolateral and medial prefrontal cortex, anterior insula, occipital cortex, and basal ganglia–thalamic circuits are also engaged, indicating a distributed control architecture. Prefrontal regions support maintenance and flexible use of task-relevant information \citep{miller2001integrative,baddeley2003working}, and cortico–striato–thalamic loops support response selection and initiation \citep{alexander1986parallel}.

\subsection{Identifying individual differences during task execution}
\label{subject}
\begin{figure}[htbp]
	\centering
	\includegraphics[scale=0.725]{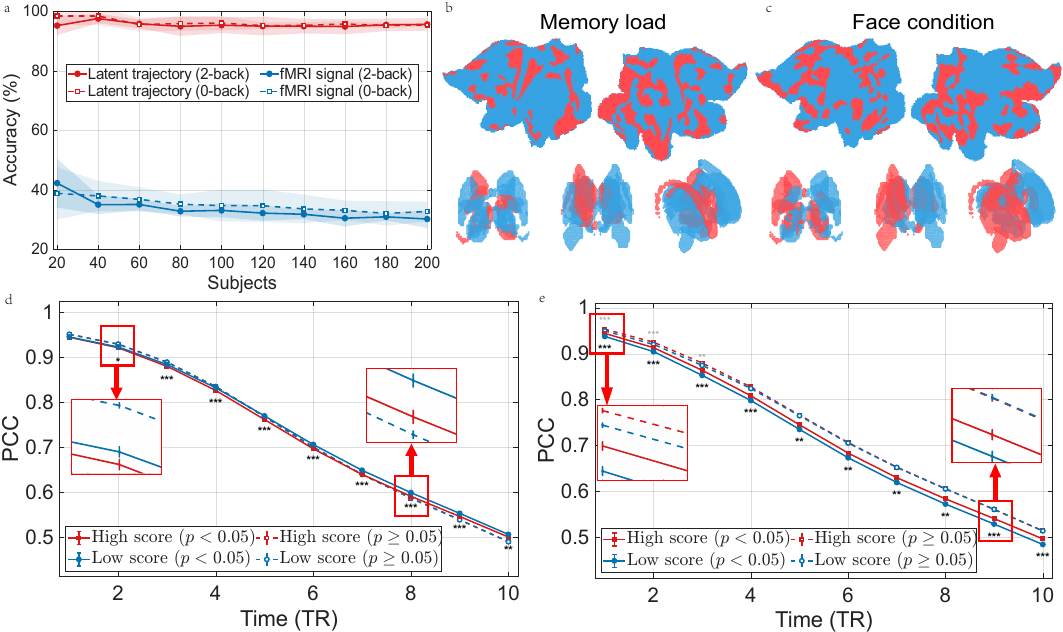}
	\caption{\textbf{Individual differences identified by brain dynamics are specific to conditions.} 
    \textbf{a}, clustering accuracy for subjects using two data types across conditions. The x-axis indicates the number of subjects (20--200) serving as clustering labels. Latent trajectories (red curves) consistently exhibit over 90\% accuracy regardless of label size, while fMRI signals (blue curves) yield consistently low accuracy. Shaded areas indicate the standard deviation across subjects. Data from the 0-back (transparent markers) and 2-back (solid markers) conditions of the working memory task show a minimal effect on the results.
    \textbf{b--c}, condition-specific individual differences across brain areas. Subjects are stratified into two groups based on their performance scores. During different conditions in the working memory task (time length=10), a Kolmogorov–Smirnov test is performed for each area to assess whether the two groups differ in brain dynamics. Red areas indicate rejection of the null hypothesis ($p<0.05$), whereas blue areas indicate failure to reject the null hypothesis ($p\geq0.05$). Panel \textbf{b} shows the results using \texttt{WM 2-back Accuracy} as the grouping variable across four blocks (face, body, place, tool). Panel \textbf{c} shows the results using \texttt{WM 2-back Face Accuracy} as the grouping variable under the face block.
    \textbf{d--e}, individual differences across two types of brain areas. Brain areas are classified into two types (denoted by solid and dashed lines) as defined in \textbf{b--c}. For each type, curves show the PCC of brain dynamics between 0-back and 2-back across 10 time points for high-score (red) and low-score (blue) groups. A one-tailed t-test is performed at each time point to compare the PCC between the two groups. Significance levels: * $p < 0.05$, ** $p < 0.01$, *** $p < 0.001$. The inset displays the standard deviation across subjects. Panel \textbf{d} corresponds to areas defined in \textbf{b}, and panel \textbf{e} corresponds to areas defined in \textbf{c}.}
    \label{pic:4}

\end{figure}

Given that brain dynamics capture underlying mechanisms across tasks and conditions, we further extended their scope to describe individual differences in behavioral performance. We first demonstrated that latent trajectories sampled from the working memory task encode more subject-specific information than raw fMRI signals by comparing their clustering accuracies (Methods). When treating each subject as a distinct label, Fig.\hyperref[pic:4]{4a} shows that latent trajectories (red curve) exhibit robust performance, maintaining an accuracy above 90\% independent of the number of subjects. In contrast, the accuracy achieved by the corresponding fMRI signals (blue curve) starts around 40\% and subsequently decreases to 30\%. These accuracies are consistent for data sampled from both the 0-back (bold markers) and 2-back (transparent markers) conditions. The results indicate that subject-specific information is richer in the universal latent space than in the original measurement space, supporting the study of individual differences through brain dynamics.

To quantify shifts in brain dynamics induced by memory load, we calculated DC between dynamics derived from the 0-back and 2-back conditions across the face, place, tool, and body blocks (time length=10). To highlight individual differences in these shifts, we stratified subjects into two groups (100 subjects per group) based on their \texttt{WM 2-back Accuracy} scores (i.e., accuracy in the 2-back condition \citep{barch2013function}) and employed the Kolmogorov-Smirnov (K-S) test to identify regional disparities (Methods).
Fig.\hyperref[pic:4]{4b} categorizes brain areas based on whether significant group differences exist ($p<0.05$, red) or not ($p\geq0.05$, blue). Focusing on these differences within distinct area types, Fig.\hyperref[pic:4]{4d} plots DC for the high-score (red) and low-score (blue) groups in two typed areas (bold and dashed lines indicate the red and blue areas, respectively). Asterisks ($*$) denote significance levels (one-sided t-test), and insets display subject variability. In red areas, the high-score group shows significantly weaker DC than the low-score group from the second time point onward, whereas no significant differences are observed in blue areas. Fig.\hyperref[pic:4]{4b} and Fig.\hyperref[pic:4]{4d} reveal a clear load-dependent pattern underlying individual differences. The higher-performing group exhibits stronger load-dependent modulation (2-back vs. 0-back) across distributed cortical and subcortical systems supporting working memory \citep{owen2005n, rottschy2012modelling}. Specifically, greater load-dependent changes are observed in visual cortices, consistent with the active maintenance of sensory representations \citep{harrison2009decoding, emrich2013distributed}. Similar effects are found in frontoparietal and dorsal attention networks, including superior parietal and prefrontal regions, which scale with load and predict capacity limits \citep{wager2003neuroimaging, todd2004capacity, vogel2005neural}. Enhanced modulation is also evident in salience and cingulo-opercular regions implicated in task-set maintenance \citep{seeley2007dissociable, dosenbach2008dual}. In addition, medial temporal regions show load-related changes that may reflect dynamic prefrontal–hippocampal interactions under increased demand \citep{rissman2008dynamic, baddeley2011working}. Furthermore, group differences in subcortical structures, including the thalamus, basal ganglia, and midbrain, suggest greater engagement of cortico-striatal circuits involved in information gating \citep{mcnab2008prefrontal}. Collectively, these findings suggest that superior N-back performance is associated with more robust, load-dependent network reconfiguration rather than uniformly greater activation. This pattern is broadly consistent with frameworks of dynamic network integration and segregation \citep{cohen2016segregation, shine2018principles}.

Building on the established link between brain dynamics and individual differences in memory load, we extended this analysis to stimulus-specific performance, focusing on face processing. Using the previously described methodology, we quantified shifts in brain dynamics restricted to the face block and stratified subjects based on \texttt{WM 2-back Face Accuracy} (i.e., accuracy during face blocks in the 2-back condition). Fig.\hyperref[pic:4]{4c} reveals that the spatial distribution of significant brain areas differs markedly from that in Fig.\hyperref[pic:4]{4b}. Fig.\hyperref[pic:4]{4e} further indicates a reversal in effect direction compared to Fig.\hyperref[pic:4]{4d}: within the red areas, the high-score group exhibits significantly stronger DC than the low-score group. In contrast, in the blue areas, this group difference diminishes over time and disappears by t=4. Integrating Fig.\hyperref[pic:4]{4c} and Fig.\hyperref[pic:4]{4e}, higher-performing individuals tend to exhibit reduced load-dependent modulation (2-back vs. 0-back) across early and higher-order visual cortices and the dorsal stream, suggesting a “task-ready” baseline requiring less upregulation under increased load \citep{kastner1999increased, corbetta2002control}. Similarly, this group shows smaller load-dependent changes in temporal face-processing areas \citep{kanwisher1997fusiform, haxby2000distributed}, insular and cingulo-opercular regions associated with salience and task-set maintenance \citep{seeley2007dissociable, dosenbach2008dual}, and medial prefrontal and cingulate regions implicated in default-mode/control network interactions \citep{spreng2010default}. This stable engagement across load conditions reflects greater neural efficiency, where visual expertise mitigates the need for additional recruitment during higher task difficulty \citep{gauthier1999activation, neubauer2009intelligence}. We further apply the same analysis to the place, tool, and body blocks (Extended Data Fig.\ref{pic:sup_wm}). Together, these results underscore that individual differences captured by brain dynamics are highly condition-specific.

\section{Discussion}

Modeling brain activity is highly challenging due to its inherent complexity and variability across cognitive states and individuals. Here, we propose a novel paradigm to address this challenge based on an intuitive concept: accurately predicting brain signals using a computational model, and subsequently quantifying the underlying neural dynamics from the model-derived representations. 
Despite its conceptual simplicity, implementing this paradigm effectively is difficult and heavily relies on high prediction accuracy. Traditional methods have struggled to achieve this, particularly at the whole-brain level \citep{poldrack2024handbook,friston2009modalities}. Although deep learning techniques have narrowed this gap, existing approaches often suffer from poor generalizability, meaning that specific experimental conditions and individual subjects still require tailored models or parameters \citep{vieira2017using,gosztolai2025marble}.
The next problem is extracting the underlying mechanisms from the established model. Mechanistic modeling typically incorporates strong prior structure, and multiple plausible mechanisms can explain similar observations, making interpretations difficult to falsify in noisy, partially observed brain signals \citep{daunizeau2011dynamic,stephan2010ten,breakspear2017dynamic}. Although purely data-driven causal discovery yields candidate graphs, these typically require interventional validation due to their sensitivity to hidden confounding, measurement error, model assumptions, and distribution shifts \citep{smith2011network,reid2019advancing,sanchez2021combining}. Hybrid approaches that combine mechanistic structure with deep learning can improve flexibility, but the learned components may remain opaque, which can limit mechanistic interpretability \citep{koppe2019identifying,luo2025mapping}.
To address the first challenge, UBD employs GCNs to integrate spatial properties through dMRI-derived SC, alongside a DKO to capture temporal increments. By leveraging these synergistic spatiotemporal properties, UBD achieves unprecedented accuracy in predicting whole-brain fMRI signals. Furthermore, even when trained on a limited number of subjects, our model generalizes its predictive accuracy across various cognitive states, broader cohorts, and different datasets.
Regarding the second challenge, the interpretable architecture of UBD allows us to extract clear mechanistic insights. Specifically, each fMRI signal corresponds to a latent trajectory whose evolution is governed by DKO-generated angular frequency, linking its temporal properties directly to frequency. We further characterize brain dynamics using the model-derived Jacobian matrix, which quantifies the whole-brain influence on the evolution of latent trajectories. Ultimately, these brain dynamics provide a versatile methodology for analyzing brain activity across diverse research objectives.  

The temporal property revealed by UBD has strongly connection to infra-slow fluctuation (ISF), which is composed of spontaneous oscillations in $0.01\sim 0.1$hz and is used to depict FC in resting-state \citep{biswal1995functional,fox2005human}. The four frequency peaks closely aligns with the classically defined fMRI frequency bands—namely, Slow-5 (0.01–0.027 Hz), Slow-4 (0.027–0.073 Hz), and Slow-3 (0.073–0.198 Hz) \citep{zuo2010oscillating} and the values of the peaks are organized in a natural logarithmic progression \citep{buzsaki2004neuronal}. Our observation that the dominant frequency band (0--0.014 Hz) is consistent with the well-established role of this ultra-slow frequency band (Slow-5) in integrating large-scale cortical networks and maintaining DMN activity during rest \citep{he2008electrophysiological}. The second frequency band (0.035--0.040 Hz) aligns with the Slow-4 band, which is predominantly associated with subcortical activity and the functional integration of sensorimotor networks \citep{baria2011anatomical}. The higher-frequency ranges (0.078–0.084 and 0.115–0.124 Hz) likely reflect systemic physiological rhythms, such as Mayer waves driven by peripheral cardiovascular activity \citep{julien2006enigma,tong2013evaluating}. To the best of our knowledge, this study provides the first empirical evidence that ISF manifests as distinct, separable spectral peaks rather than a continuous distribution. Conceptually, this discrete ultra-slow BOLD architecture mirrors established electrophysiological findings, where distinct functional brain networks are similarly governed by specific frequency signatures of neural oscillations \citep{mantini2007electrophysiological,chan2015mesoscale}.

As a highly complex network system, the brain relies on the intricate collaboration between static physical wiring (i.e., SC) and dynamic neural synchronization (i.e., FC) \citep{bullmore2009complex,sporns2013structure,park2013structural}.  In recent years, researchers have increasingly recognized that understanding the mechanisms of structural-functional coupling (SFC)—specifically, how functional networks dynamically emerge from the anatomical scaffold and undergo a certain degree of dissociation (decoupling)—is central to uncovering the neural basis of complex cognition \citep{suarez2020linking,baum2020development}. Crucially, UBD precisely predicts fMRI signals by incorporating time delays and dMRI-derived structural constraints \citep{deco2011emerging,mitra2015lag}, and the model-derived brain dynamics explicitly capture the temporal evolution from anatomical scaffolding to functional emergence. In the initial time, brain activity is predominantly governed by direct anatomical connections, reflecting first-order signal transmission along the structural wiring \citep{honey2009predicting,goni2014resting}. As time progresses, network communication unfolds through indirect, high-order pathways \citep{meier2016mapping,avena2018communication}.
This progression computationally validates that realistic FC is not a mere mirror of direct SC, but rather an emergent property shaped by multi-synaptic interactions over time, consistent with the recently proposed "structural-functional decoupling" hypothesis \citep{preti2017dynamic,vazquez2019gradients}.

Extracting task-related brain activity information from fMRI data typically involves constructing features such as regional activation patterns between brain areas \citep{cole2014intrinsic,haynes2015primer}, which can then be analyzed using statistical or machine learning approaches to identify and decode different task-related brain states \citep{shine2016dynamics,bessadok2022graph}. 
Here, we introduce a novel approach to accomplish this goal by leveraging the model-derived Jacobian matrix, which directly reflects the underlying continuous dynamics of brain activity. Specifically, we extract task-related information by comparing Jacobian matrices derived from task onset and offset periods. Building upon the robust predictive performance of our model, this analytical framework can be generalized across various tasks, thereby capturing task-related neural dynamics at a finer temporal granularity. Different cognitive task states produce distinct spatial patterns of brain activity; even within a single paradigm, neural dynamics are hierarchically organized and continually reconfigure as the task unfolds \citep{cohen2016segregation,vidaurre2017brain}. Extending this perspective, UBD effectively disentangles distinct brain activities within a single task block. Crucially, these findings indicate that transitions between brain states are driven by specific cognitive demands and processing stages, rather than mere chronological progression \citep{gonzalez2015tracking,hayden2025rethinking}.

Individual differences are central to neuroscience but are often confounded by measurement noise and methodological variability \citep{dubois2016building,hedge2018reliability}. Traditional activation-based and connectivity-based methods, alongside emerging approaches like connectome-based predictive modeling (CPM), aim to characterize these variations by linking brain measures to performance scores across individuals \citep{finn2015functional,shen2017using}. However, the stark dimensional contrast between brain features to behavioral score (e.g., accuracy or reaction time) and to sample size (i.e., the p$\gg$n problem) makes model estimation inherently challenging \citep{bzdok2018machine,poldrack2020establishment,liu2023replicable}. While leave-one-out cross-validation (LOOCV) is frequently employed to improve model estimation, its evaluation is still constrained by high variance in performance estimates and potential optimistic bias \citep{varoquaux2017assessing}.
Since subject-specific information is complex, relying solely on fitting behavioral scores is often insufficient to capture it comprehensively. To circumvent this limitation, we characterize each subject using latent trajectories, then use the Jacobian matrix to manifest individual information in brain measures. After identifying individual differences, we explain these differences by comparing subject groups stratified by their behavioral performance. Our results suggest individual differences during tasks reflect network-wide properties rather than localized activity in specific regions \citep{rosen2025distributed}. Crucially, our proposed method offers a more flexible framework for connecting brain measures to individual traits. Although we utilize a single performance score to define behavioral differences in this study, the approach can easily accommodate multidimensional metrics (e.g., the g-factor \citep{smith2015positive}). Furthermore, while our current analysis centers on the working memory task, the analytical framework is readily generalizable to other cognitive tasks. 

Finally, the proposed paradigm holds significant potential for broader application across diverse neuroscientific and clinical domains. Its inherent flexibility allows it to accommodate various data modalities (e.g., EEG) and adapt to different ROI templates, thereby enabling the analysis of brain dynamics across broad spatial and temporal scales. Given its success in characterizing individual differences, the paradigm offers a promising avenue for discovering novel disease biomarkers through patient-control comparisons. In clinical settings, the capacity of our paradigm to analyze single subjects on a second-by-second basis facilitates the extraction of highly individualized diagnostic information.

\section{Methods}

\subsection{Data pre-process}

For resting-state fMRI data, each participant underwent resting-state fMRI over two consecutive days, with two 15-minute runs per session (4 runs total), collected at 2 mm isotropic resolution and a TR of 720 ms, consistent with HCP acquisition protocols. Task-based fMRI followed the same acquisition scheme. All data were preprocessed using the HCP minimal pipeline (FSL, FreeSurfer, FIX-based ICA denoising), including correction for spatial and gradient distortions, motion correction, registration to T1-weighted structural scans, and normalization to standard space \citep{glasser2013minimal}. Finally, the voxel-wise data were parcellated using the HCPex atlas, yielding time series for 426 distinct brain regions \citep{huang2022extended}. A Butterworth filter was applied to retain data in the 0.01–0.1 Hz range. A snapshot $\bm{x} \in \mathbb{R}^{426\times d}$ was stacked by time-delay embedding, where $d$ is the time-delay embedding length.

The dMRI data were acquired on a Siemens 3T Skyra scanner using a 2D spin-echo single-shot multiband EPI sequence with a multiband factor of 3 and monopolar gradient pulse. The data consisted of three shells (b-values: 1000, 2000, and 3000 s/mm$^2$), with 90 diffusion directions equally distributed within each shell (270 directions in total), plus 6 b=0 acquisitions per shell, at a spatial resolution of 1.25 mm isotropic voxels. Data were preprocessed with the HCP diffusion preprocessing pipeline (\href{https://www.humanconnectome.org/study/hcp-young-adult/data-releases}{v3.19.0}), which consisted of b0 intensity normalization across runs, EPI distortion correction, eddy-current-induced distortion correction, motion correction, gradient nonlinearity correction, registration to native structural space, and masking the final data with a brain mask \citep{glasser2013minimal}. Subsequently, a probabilistic tractography pipeline implemented in MRtrix3 was performed to generate streamlines \citep{tournier2019mrtrix3}, utilizing Multi-Shell Multi-Tissue Constrained Spherical Deconvolution (MSMT-CSD) to resolve fiber orientations \citep{jeurissen2014multi} and the iFOD2 algorithm to perform whole-brain tracking \citep{tournier2010improved}. To construct the individual SC matrices, the standard-space streamlines were mapped to the same 426 brain regions defined by the HCPex atlas \citep{huang2022extended}. Finally, a group-averaged SC across 161 subjects was computed.

Although weighted SC conveys detailed streamline counts representing connection strength, we intentionally binarized the group-averaged SC using a minimal threshold (value $\ge 1$ set to 1) to establish a purely topological scaffold. The resulting group SC contained 39,991 non-zero elements (181,476 in total), and no isolated vertices were present. This design isolates the role of physical routing from connectivity magnitude, forcing the GCN to learn dynamic functional hierarchies solely constrained by anatomical presence.

\subsection{Model architecture}
The model comprises three ingredients: the encoder $\varphi$, the decoder $\phi$ and DKO $\mathcal{A}$ \citep{luschDeepLearningUniversal2018}. The encoder and the decoder are implemented by GCN \citep{fey2019fast,morris2019weisfeiler}. Specifically, $L$-layers of the encoder are guided by the routing contained in SC to generate the latent representation of $\bm{h}^{l}=\varphi(\bm{h}^{l})$ as
\begin{align}
    \bm{h}^{l+1}_{i}=\bm{w}_{1}\bm{h}^{l}_{i}+\bm{w}_{2}\sum_{j\in \mathcal{N} (i)}e_{j,i} \bm{h}^{l}_{j}
\end{align}
where \(e_{j,i}\) denotes the connection from source node \(j\) to target node \(i\), and \(\bm{w}_1,\bm{w}_2\) are learned from data. Specifically, \(\bm{h}^{0}=\bm{x}\) is a snapshot of fMRI signals, and \(\bm{h}^{L}\) is the corresponding latent representation. For \(\bm{h}\in\mathbb{R}^{426\times d}\), the first dimension corresponds to the number of brain areas, and the second is the time-delay embedding length. The decoder $\phi$ maps \(\bm{h}^{L}\) back to the measurement space via \(\bm{h}^{0}=\phi(\bm{h}^{L})\), mirroring the encoder via the identical SC.

To learn the temporal increments of $\bm{h}(t)$ in the latent space, $\mathcal{A}$ employs an MLP to generate $\bm{\theta}(t)$ from $\bm{h}(t)$ at each time point. The components of $\bm{h}(t)$ are arranged in complex conjugate pairs, with each pair sharing the same angular frequency $\theta_{i}(t)$ (collected in $\bm{\theta}(t)$). The update from $t$ to $t+1$ is implemented as an element-wise complex rotation
\begin{align}
    \bm{h}(t+1)=\bm{h}(t)\odot e^{i\bm{\theta}(t)\Delta t},
\end{align}
which corresponds to multiplying each complex component by a unit-magnitude complex exponential. Because the fMRI signal is sampled every 0.72 seconds, we set $\Delta t = 0.72$. The latent trajectory is driven as $\bm{h}(t)= \underbrace{\mathcal{A}\circ\mathcal{A}\circ\cdots\circ\mathcal{A}}_{t} (\bm{h}(0))$, we mark as $\bm{h}(t)=\mathcal{A}^{t} \circ \bm{h}(0)$.

\subsection{Model training and prediction}
To train the model, we collected resting-state fMRI data from 35 subjects and discarded the first and last 200 time points, retaining 800 time points per subject. We then segmented each subject’s data into multiple snapshots and pooled all snapshots across subjects to form the training dataset. To update our model, we use temporal snapshot sequences \(\bm{X} \in \mathbb{R}^{T\times 426\times d}\) (set $T$ to 10 and $d$ to 18 in this study) to derive prediction loss and latent loss. Prediction loss computes the mean square error (MSE) between predicted and observed fMRI snapshots as
\begin{align}
    \sum_{t=2}^{t=T}\lVert \bm{X}(t) - \phi \circ\mathcal{A}^{t}\circ\varphi \circ \bm{X}(0) \rVert_{2}^{2}.
\end{align}
Latent loss computes the MSE between the latent trajectory evolved from the initial time via $\mathcal{A}$, and mapped at each time via the encoder $\varphi$ as 
\begin{align}
    \sum_{t=2}^{t=T}\lVert \varphi \circ \bm{X}(t) - \mathcal{A}^{t}\circ\varphi \circ \bm{X}(0) \rVert_{2}^{2}.
\end{align}
The model was trained for 700 epochs using the Adam optimizer \citep{kingma2014adam} with a learning rate of $10^{-4}$.

The trained model predicts fMRI data using the first \(d\) time points \(\bm{x}(0)\in \mathbb{R}^{426\times d}\) as input
\begin{align}
\bm{\hat{\bm{X}}}(\hat{t})=\phi\circ\mathcal{A}^{\hat{t}}\circ\varphi\circ\bm{x}(0),
\end{align}
where $\bm{\hat{\bm{X}}}(\hat{t}) \in \mathbb{R}^{\hat{t}\times 426\times d}$, \(\hat{t}\) is the prediction horizon. We obtain the prediction results $\bm{\hat{x}}\in\mathbb{R}^{\hat{t}\times 426}$ by extracting the first element along the third dimension of \(\bm{\hat{X}}\). We compute the PCC between prediction $\bm{\hat{x}}$ and observation $\bm{x}$ for each brain area over time like
\begin{align}
    \bm{p}_{t,m}=\mathrm{corr}(\bm{\hat{x}}(1:t,m), \bm{x}(1:t,m)), \quad t=1,\dots,\hat{t}, \quad m=1,\dots,426
\end{align}
where \(\bm{p}\in \mathbb{R}^{\hat{t} \times 426}\), $m$ corresponds to a specific brain area. Fig.\hyperref[pic:1]{1c} shows the values of \(\bm{p}\) when $\hat{t}=100$, while Fig.\hyperref[pic:1]{1b} plots a single row representing the prediction results for one area.

\subsection{Angular frequency to frequency conversion}
The latent space is spanned by basis functions learned by the encoder and decoder, analogous to the Fourier basis in the frequency domain. The evolution of the latent trajectory $\bm{h}(t+1)=\bm{h}(t)\odot e^{i\bm{\theta}(t)\Delta t}$ corresponds to an element-wise phase rotation, where $\bm{\theta}(t)\Delta t$ represents the phase increment over one time step. Since $\Delta t=0.72\,\mathrm{s}$, $\bm{\theta}(t)$ has units of rad/s. The corresponding instantaneous frequency is then given by
\begin{align}
    \bm{f}(t)=\frac{\bm{\theta}(t)}{2\pi},
\end{align}
where $\bm{f}(t)$ is measured in Hz.

\subsection{Kruskal-Wallis test of angular frequency distribution}

To assess whether the $\bm{\theta}(t)$ distributions differed significantly across subjects, we performed the Kruskal-Wallis H-test (one-way ANOVA on ranks, implemented by MATLAB function \texttt{kruskalwallis}). This non-parametric test was chosen to evaluate the null hypothesis that data samples from all subjects originate from the same distribution, without assuming a normal distribution of the underlying data. The subject identity served as the independent grouping variable. Considering that statistical significance (P-value) can be easily inflated by large sample sizes, we complemented the hypothesis testing by calculating the effect size, Epsilon-squared ($\epsilon^{2}$). This metric was derived from the H-statistic ($\chi^{2}$) to quantify the proportion of variance in the feature data that can be attributed to subject identity, thereby providing a measure of the magnitude of inter-subject discriminability independent of sample size.

\subsection{Quantification of brain dynamics via the Jacobian matrix}
The Jacobian matrix is computed using automatic differentiation in PyTorch \citep{paszke2019pytorch}. We focus on the influence of each brain area, which mainly relates to the encoder $\varphi$ and the auxiliary $\mathcal{A}$. Given the initial fMRI snapshot $\bm{x}(0)\in \mathbb{R}^{426\times d}$ in the measurement space, our model represents brain activity at time $t$ as
\begin{align}
    \bm{h}(t) = \mathcal{A}^{t} \circ \varphi \circ \bm{x}(0), \quad \bm{h}(t) \in \mathbb{R}^{426\times d}.
\end{align}
While $\bm{h}(t)$ and $\bm{x}(0)$ have the same dimension, the first dimension does not represent a brain area because $\bm{h}(t)$ resides in the latent space. We use the Jacobian matrix to quantify how input brain areas influence the latent trajectory at time $t$:
\begin{align}
    \mathcal{\bm{J}}(t) = \frac{\partial \bm{h}(t)}{\partial \bm{x}(0)}, \quad \mathcal{J}(t) \in \mathbb{R}^{426 \times d \times 426 \times d}.
\end{align}
While strictly topological perturbations in delay-embedded spaces should primarily propagate from the leading temporal edge, we averaged the Jacobian matrix over the second and fourth dimensions (i.e., the time-delay embedding length $d$). This simplification is intentionally designed to capture the generalized, time-integrated susceptibility of regional dynamics rather than instantaneous state-space perturbations, yielding a tractable macroscopic brain dynamics matrix $\bm{j}(t) \in \mathbb{R}^{426 \times 426}$. Each column of $\bm{j}(t)$ represents the integral influence of one input brain area on $\bm{h}(t)$, thereby linking the macroscopic measurement space to the latent trajectory.

Since each column of the brain dynamics matrix $\bm{j}(t)$ represents the influence profile of a single brain area, we compute the pairwise PCC between columns to quantify similarity among brain areas, resulting in a correlation matrix $\mathrm{corr}(t) \in \mathbb{R}^{426 \times 426}$. For each brain area, we compute the PCC between $\mathrm{corr}(t)$ and SC/FC, then average across areas to obtain individual results for 963 subjects. For SC, we correlate each subject’s $\mathrm{corr}(t)$ with the SC used to train UBD; for FC, we correlate each subject’s $\mathrm{corr}(t)$ with their own FC. In section \ref{rest}, Fig.\hyperref[pic:2]{2d} and Fig.\hyperref[pic:2]{2f} illustrate $\mathrm{corr}(t)$ in resting-state fMRI signals at times $t=1$ and $t=10$, respectively.

To quantify the changes driving cognitive transitions, we derive brain dynamics for two cognitive conditions, denoted by $\bm{j}_a$ and $\bm{j}_b$. To measure the change in the influence of each brain area, we compute the PCC between the corresponding columns of $\bm{j}_a$ and $\bm{j}_b$, yielding $\mathrm{corr}_{a,b} \in \mathbb{R}^{426 \times 1}$. We refer to this measure as the dynamics correlation (DC). In section \ref{task}, Brain dynamics were derived from 915 subjects in the motor task and 920 in the language task, and DC was computed using the group means of their absolute values.

\subsection{Cognitive state classification}

The classification task aims to distinguish fMRI data acquired under different task-evoked cognitive states (including resting-state, working memory, language, motor, social cognition, relational processing, emotion processing, and gambling) using supervised learning. For each subject, the fMRI series was denoted as $\bm{x}\in\mathbb{R}^{t\times426}$, where $t$ is the number of time points and denotes the x-axis of Fig.\hyperref[pic:3]{3a}. We used time-delay embedding to generate a series of snapshots $\bm{X}\in\mathbb{R}^{(t-d)\times 426 \times d}$, and used the encoder $\varphi$ to generate the corresponding latent representations $\bm{H}=\varphi(\bm{X})\in\mathbb{R}^{(t-d)\times 426 \times d}$. To highlight the effective of $\varphi$, we removed temporal ordering by shuffling the first dimension of $\bm{X}$ and $\bm{H}$. Data were collected from 50 subjects across eight states, and all snapshots were pooled to form the dataset $\mathcal{D} \in \mathbb{R}^{8 \times 50 \times (t-d) \times 426 \times d}$. Each sample $\bm{d} \in  \mathbb{R}^{ 426 \times d}$ in $\mathcal{D}$ was assigned one of the eight state labels. A multiclass Support Vector Machine (SVM) with a Gaussian kernel was implemented using the MATLAB function \texttt{fitcecoc} to classify eight distinct states. The dataset was divided into $70\%$ training and $30\%$ testing sets using stratified hold-out validation with a fixed random seed for reproducibility. Classification accuracy was calculated as the percentage of correctly predicted samples in the test set as
\begin{align}
    \mathrm{Accuracy}=\frac{\sum \mathrm{I}(y_{\mathrm{pred}}=y_{\mathrm{true}})}{N_{\mathrm{test}}},
\end{align}
where $\mathrm{I}$ is the indicator function and $N_{\mathrm{test}}$ is the total number of test samples.

\subsection{Subject identification via clustering}

The clustering task aims to distinguish fMRI data from different subjects using unsupervised learning, where each subject is treated as a distinct cluster label. For each subject, we first construct a sequence of fMRI snapshots $\bm{X} \in \mathbb{R}^{t\times 426\times d}$ sampled from the working memory task. Starting from the initial snapshot $\bm{x}(0)$, the corresponding latent trajectory is generated through
\begin{align}
    \bm{H} =\mathcal{A}^{t} \circ \varphi \circ  \bm{x}(0)\in \mathbb{R}^{t\times 426\times d}.
\end{align}
We collect sequences of 25 time points from 4 blocks (face, body, tool, place). All sequences were then pooled to form the dataset $\mathcal{D} \in \mathbb{R}^{n \times 4 \times 25 \times 426\times d}$, where $n$ denotes the number of subjects and corresponds to the x-axis in Fig.\hyperref[pic:4]{4a}. Thus, each sample $\bm{d} \in \mathbb{R}^{25 \times 426 \times d}$ in $\mathcal{D}$ represents a sequence associated with a specific subject and task execution, while the subject identity serves as the ground-truth label for evaluation.

To group the sequences by subject identity, we applied spectral clustering using the MATLAB function \texttt{spectralcluster}, with the number of clusters set to the number of subjects. When the number of clusters increased to around 50, numerical issues arose when operating on sequences of fMRI snapshots (but not on latent trajectories). In this case, we instead used the \texttt{knn} function and report the better performance achieved by either \texttt{spectralcluster} or \texttt{knn}. Since clustering outputs arbitrary cluster indices, the Hungarian algorithm \citep{kuhn1955hungarian} was further employed to obtain the optimal assignment between predicted cluster labels and the ground-truth subject labels for evaluation. The same procedure is repeated for sequences sampled from the 0-back and 2-back conditions.

\subsection{Identifying individual differences via dynamics correlation}

Dynamics correlation (DC) quantifies the changes in brain dynamics associated with cognitive transitions. For a time series, DC is computed as $\mathrm{corr} \in \mathbb{R}^{t\times 426}$, where $t$ denotes the number of time points and 426 represents the brain regions. Subjects were divided into high-score and low-score groups, each containing $n$ subjects. DC values for the two groups were organized as $\mathrm{corr}_{h}, \mathrm{corr}_{l} \in \mathbb{R}^{n\times t\times 426}$.

To identify brain regions exhibiting significant differences between the two groups, we performed a two-sample Kolmogorov–Smirnov (KS) test using the MATLAB function \texttt{kstest2}. The test was conducted independently for each brain area, comparing the distributions of $\mathrm{corr}_{h}$ and $\mathrm{corr}_{l}$ across subjects and time points. This non-parametric test evaluates the null hypothesis that samples from the high-score and low-score groups are drawn from the same continuous distribution. According to whether the null hypothesis is rejected, we define a binary significance vector $\mathbf{s}$ as
\begin{align}
    s_i =
    \begin{cases}
    1, & p_i < 0.05, \\
    0, & p_i \geq 0.05,
    \end{cases}
    \quad i = 1, \dots, 426,
\end{align}
where each entry indicates whether the corresponding brain region shows a statistically significant difference between the two groups. For visualization, entries with values of 1 and 0 in $\mathbf{s}$ were colored red and blue, respectively, as shown in Fig.\hyperref[pic:4]{4b} and Fig.\hyperref[pic:4]{4c}.

We used this procedure to examine individual differences in memory load and face processing within the working memory task ($n=100$, $t=10$). For memory load, DC is sampled from 4 blocks (face, body, tool, place), yielding $\mathrm{corr}_{h}, \mathrm{corr}_{l} \in \mathbb{R}^{100\times 40\times 426}$. Subjects were stratified into high-score and low-score groups based on \texttt{WM 2-back Accuracy}. For the face processing, DC is sampled only from the face block, yielding $\mathrm{corr}_{h}, \mathrm{corr}_{l} \in \mathbb{R}^{100\times 10\times 426}$, where group stratification was determined by \texttt{WM 2-back Face Accuracy}.

\subsection{Data availability}
The HCP-YA data used in this study are available in the Human Connectome Project database under the accession link \url{https://www.humanconnectome.org/study/hcp-young-adult/data-releases}. The HCP-YA data are available under restricted access for compliance with data use policies.

\subsection{Code availability}
The code for the main results of this paper is provided in \url{https://github.com/Rrh-Zheng/Universal-Brain-Dynamics}.

\subsection*{Acknowledgement}
This work was partially supported by the National Natural Science Foundation of China (under grant No. 12471481) and the Science and Technology Commission of Shanghai Municipality (under grant No. 23ZR1403000).

\newpage
\bibliography{reference}

\newpage

\renewcommand{\figurename}{Extended Data Fig.}
\setcounter{figure}{0}

\begin{figure}
	\centering
	\includegraphics[scale=0.5]{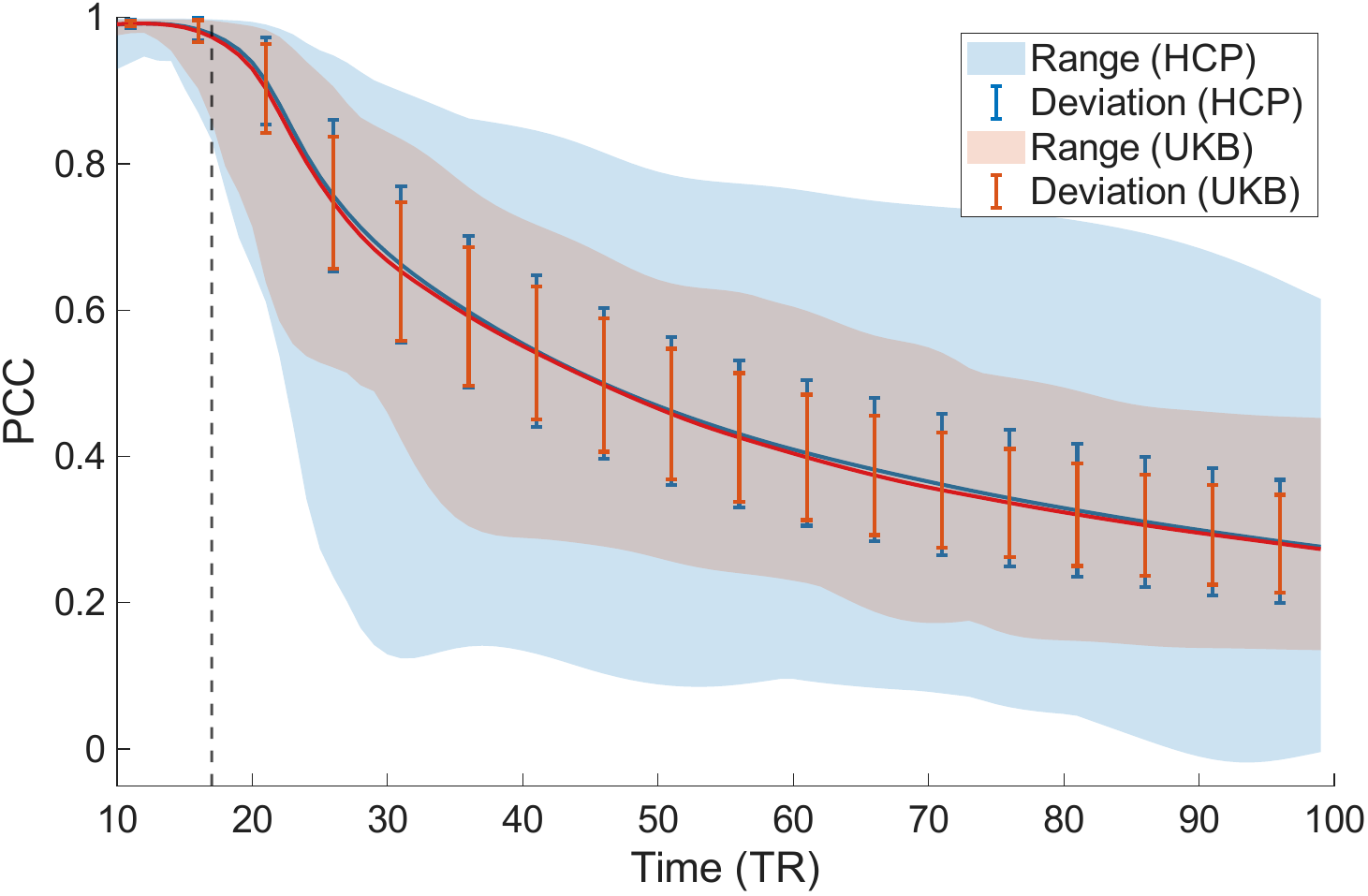}
	\caption{\textbf{Comparison of resting-state fMRI predictive performance between UKB and HCP.} 
    UBD is trained on resting-state fMRI data from 35 HCP subjects. The x-axis denotes the predictive horizon ($t=100$). The red curve represents the mean PCC between observations and predictions across 100 UKB subjects, while the blue curve represents the mean across 963 HCP subjects. Shaded areas represent the range from best to worst single-subject performance, and error bars denote the standard deviation across subjects.}
    \label{pic:sup_ukb}
\end{figure}

\begin{figure}
	\centering
	\includegraphics[scale=0.41]{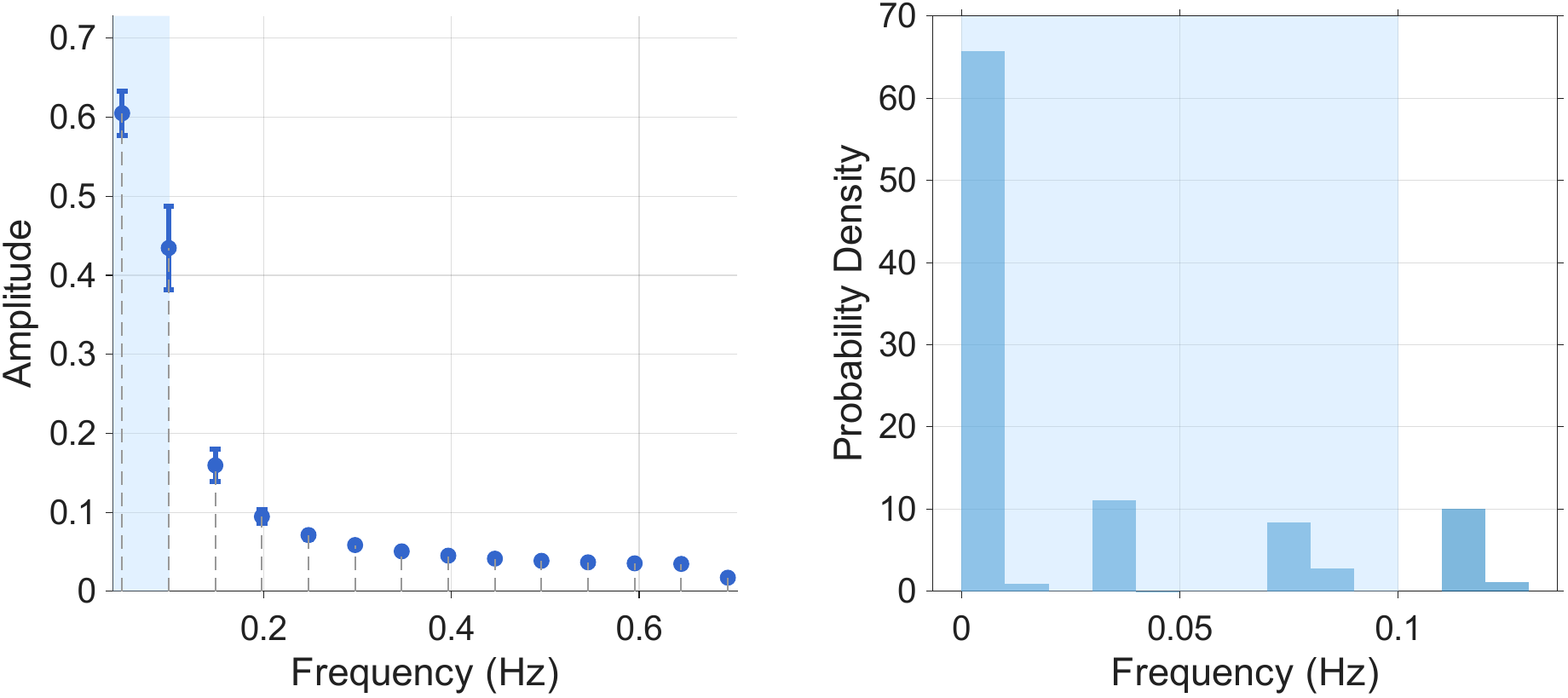}
	\caption{\textbf{The amplitude spectrum and corresponding $\bm{\theta}$ of fMRI signals.} 
    The right panel shows the amplitude spectrum of resting-state fMRI signals from 963 subjects, with error bars indicating the standard deviation across subjects. For each subject and brain area, preprocessed time series were detrended and transformed into the frequency domain using the fast Fourier transform (FFT), from which the single-sided amplitude spectrum was obtained. Amplitude spectra were averaged across brain regions to yield subject-level spectra. Although a band-pass filter (0.01–-0.1 Hz) was applied, small residual amplitudes are still observable above 0.1 Hz. This is expected due to the non-ideal frequency response of practical filters and spectral leakage induced by the short time window (time length=28, which is consistent with the estimation of $\bm{\theta}$). Therefore, the resulting spectra provide a coarse characterization of frequency content and are interpreted as relative fluctuation profiles rather than precise estimates of canonical low-frequency oscillations. The left panel shows the distribution of $\bm{\theta}$ in the frequency domain. The translucent blue region indicates the 0–0.1 Hz frequency band.}
    \label{pic:sup_spectrum}
\end{figure}

\begin{figure}[htbp]
	\centering
	\includegraphics[scale=0.73]{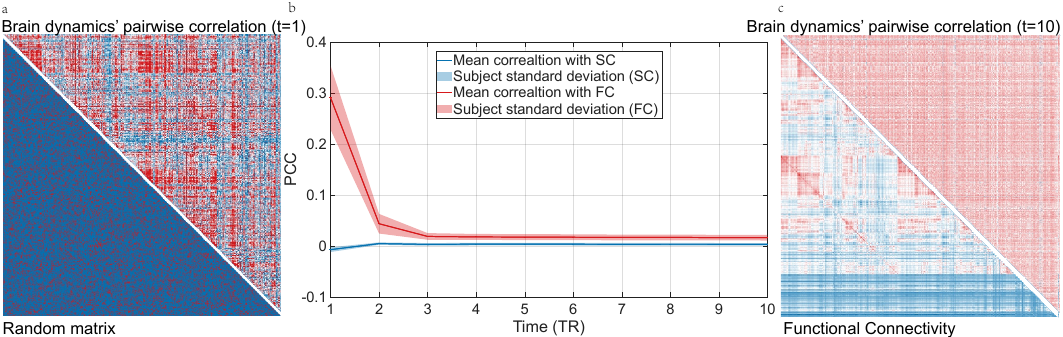}
	\caption{\textbf{Derive brain dynamics from UBD integrated within a random matrix.} During training and brain dynamics derivation, the SC within GCNs is replaced by a random matrix that preserves the same number of non-zero elements. \textbf{a--c} were generated using the same procedure as Fig.\hyperref[pic:2]{2d--e}.}
    \label{pic:sup_random}
\end{figure}

\begin{figure}[htbp]
	\centering
	\includegraphics[scale=0.73]{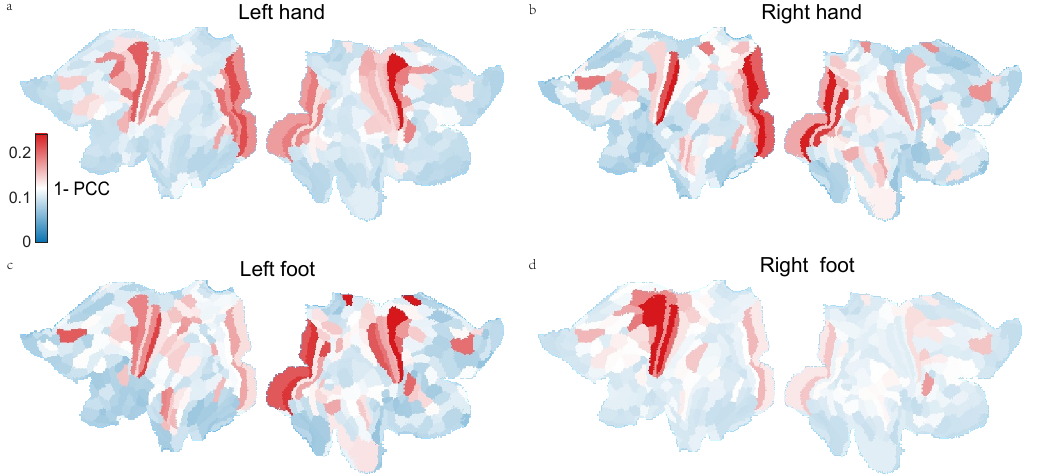}
	\caption{\textbf{Divergence of brain dynamics driven by specific movements in the motor task.} Following Fig.\hyperref[pic:3]{2b}, the panels display the dissimilarity in brain dynamics between the fixation and four movement conditions: left hand (\textbf{a}), right hand (\textbf{b}), left foot (\textbf{c}), and right foot (\textbf{d}).}
    \label{pic:sup_mot}
\end{figure}

\begin{figure}[htbp]
	\centering
	\includegraphics[scale=0.73]{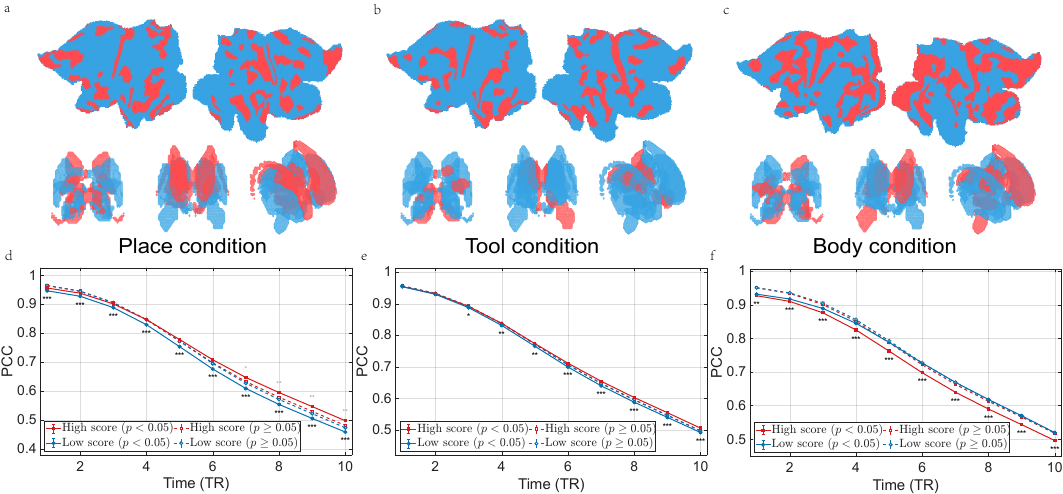}
	\caption{\textbf{Individual differences across specific conditions in the working memory task.} The generation procedures for \textbf{a--c} and \textbf{d--f} follow Fig.\hyperref[pic:4]{4c} and Fig.\hyperref[pic:4]{4e}, respectively. Both grouping variables and derived brain dynamics are condition-specific: place (\texttt{WM 2-back Place Accuracy}) for \textbf{a,d}, tool (\texttt{WM 2-back Tool Accuracy}) for \textbf{b,e}, and body (\texttt{WM 2-back Body Accuracy}) for \textbf{c,f}. Significance levels: * $p < 0.05$, ** $p < 0.01$, *** $p < 0.001$, based on a right-tailed t-test for \textbf{d, e} and a left-tailed t-test for \textbf{f}. }
    \label{pic:sup_wm}
\end{figure}

\end{document}